\documentclass[preprint,times]{emulateapj}
\usepackage{natbib}

\usepackage{graphicx}
\usepackage{epsf}
\usepackage{color} 
\usepackage{amssymb}

\begin{document}

\title{Discovery of Carbon Monoxide in Distant Comet C/2017 K2 (PANSTARRS)}
\author{Bin Yang$^{1}$, David Jewitt$^{2,3}$, Yuhui Zhao$^{4}$, Xuejian Jiang$^{5}$, Quanzhi Ye$^{6}$, Ying-Tung Chen$^{7}$ \\}

\affil{$1$ European Southern Observatory, Alonso de Córdova 3107, Santiago, Chile\\}
\affil{$2$ Department of Earth, Planetary and Space Sciences, University of California at Los Angeles, 595 Charles E Young Dr E, Los Angeles, CA 90095, USA \\ } 
\affil{$3$ Department of Physics and Astronomy, University of California at Los Angeles, 475 Portola Plaza, Los Angeles, CA 90095, USA \\ } 
\affil{$4$ CAS Key Laboratory of Planetary Sciences, Purple Mountain Observatory, Chinese Academy of Sciences, 210008 Nanjing, China; CAS Center for Excellence in Comparative Planetology, Chinese Academy of Sciences, 230026, Hefei, China \\} 
\affil{$5$ East Asian Observatory, 660 N. A`oh\={o}k\={u} Place, Hilo, HI 96720, USA \\ } 
\affil{$6$ Department of Astronomy, University of Maryland, College Park, MD 20742, USA\\}
\affil{$7$ Institute of Astronomy and Astrophysics, Academia Sinica, Taipei 10617, Taiwan}
\email{byang@eso.org}

\begin{abstract}
Optical observations of the Oort cloud comet C/2017 K2 (PANSTARRS) show that its activity began at large heliocentric distances (up to 35 au), which cannot be explained by either the sublimation or the crystallization of water ice. Supervolatile sublimation, most likely of carbon monoxide (CO), has been proposed as a plausible driver of the observed mass loss. Here, we present the detection of the J = 2$-$1 rotational transition in outgassed CO from C/2017 K2 when at heliocentric distance $r_H$ = 6.72 au, using the James Clerk Maxwell Telescope. The CO line is blue-shifted by 0.20$\pm$0.03 km s$^{-1}$ with an area and width of 8.3$\pm$2.3 mK km s$^{-1}$ and $0.28\pm$0.08 km s$^{-1}$, respectively. The CO production rate is $Q_{CO} = (1.6\pm0.5) \times10^{27}$ s$^{-1}$. These are the first observations of a gaseous species in C/2017 K2 and provide observational confirmation of the role of supervolatile sublimation in this comet.

  \end{abstract}


\section{Introduction}
Comets are primitive objects, composed of refractory solids and ices, which are well-preserved relic planetesimals since the early Solar system. In the heat of the Sun, comets  develop comae and/or tails when entering the inner Solar system. Most active comets are detected within the orbit of Jupiter, where the temperature is high enough for water ice, the major cometary volatile, to sublimate. However, a number of comets exhibit distant activity well beyond 5 au. The mechanism for this distant activity is not clear. Suggested scenarios include, the sublimation of supervolatiles (e.g.~CO and CO$_2$), the crystallization of amorphous water ice \citep{Prialnik2004}, and chemical reactions involving unstable radicals created by cosmic ray bombardment \citep{Donn:1956,Rettig:1992}. As pointed out by \cite{Roemer:1962}, our physical understanding of comets mostly comes from observing cometary activity at moderately small heliocentric distances within 5 au. The characterization of distant comets, though challenging, provides important information about the formation conditions of comets in the Solar protoplanetary disk, such as the dynamics, collisions, physical and chemical processes \citep{Meech:2001}.   

C/2017 K2 (PANSTARRS) (hereafter K2) is an Oort cloud comet on its way to perihelion, expected at heliocentric distance $r_H$ = 1.80 au in late 2022. It was discovered at $r_H$ = 16 au \citep{Wainscoat:2017}, then later identified in prediscovery images at $r_H$ = 24 au \citep{Jewitt:2017, Hui:2018}, and activity is inferred to have begun at $r_H \sim$ 35 au \citep{Jewitt:2021}. The early detection of K2 when in the outer solar system offers the unprecedented opportunity to examine the development of  activity in a comet arriving from near-interstellar temperatures. \\

The barycentric semimajor axis, eccentricity and inclination of K2 are $a$ = 20,000 au, $e$ = 0.9998 and $i$ = 87.5 deg, respectively \citep{Krolikowska:2018}. These orbital parameters qualify K2 as a long-period (Oort cloud) comet, albeit one that is not strictly dynamically new. Nevertheless, with orbit period $\sim$3 Myr, the previous perihelion occurred so long ago that the nucleus can retain no heat, and it is appropriate to consider K2 as arriving from both Oort cloud distances (aphelion is at $\sim$40,000 au) and temperatures ($\lesssim$ 10 K). The observed activity is a response to contemporaneous heating by the Sun.   \\

Optical observations of the dust coma reveal a number of distinguishing characteristics. The coma is massive, with an appearance dominated by large (submillimeter and larger) particles  ejected slowly (speeds $\sim$ 4 m s$^{-1}$ \citep{Jewitt:2017,Jewitt:2019,Jewitt:2021,Hui:2018} even in the earliest observations. The dust production is in steady state (on the coma residence timescale) but rises as the heliocentric distance falls, in proportion to $r_H^{-2}$, with an estimated value 10$^3(a_1)^{1/2}$ kg s$^{-1}$ at 10 au, where $0.1 \lesssim a_1 \lesssim 1$ is the mean dust particle radius in millimeters \citep{Jewitt:2021}.  \\

The rotational lines of CO are intrinsically weak due to the small dipole moment of this molecule. Nevertheless, the 230.5 GHz CO J(2-1) line has been detected in several comets at distances too large, and temperatures too low, for the activity to be caused by the thermodynamics of water ice. For example, CO is strong in 29P/Schwassmann-Wachmann 1, which has a nearly circular orbit at $r_H \sim$ 6 au \citep{Senay:1994}. The CO line was discovered in comet C/1995 O1 (Hale-Bopp) when inbound at $r_H$ = 6.76 au \citep{Jewitt:1996} and was followed out to $r_H$ = 14 au \citep{biver:2002}. For this reason, and given the substantial activity indicated by the optical observations, we undertook to search for rotational emission lines from CO in comet K2.\\

\section{Observations and Data Reduction}
Observations were taken at the James Clerk Maxwell Telescope (JCMT) on Mauna Kea, Hawaii. We obtained $\sim$ 8 hours of integration on K2. A journal of these observations is provided in Table \ref{obstable}. We used  the newly commissioned heterodyne instrument N\={a}makanui with its 230-GHz \`{}\={U}\`{}\={u} receiver. The half-power beam-width of \`{}\={U}\`{}\={u} is 20$\farcs$41 at 230 GHz, while the main-beam efficiency $\eta_{MB}$ is about 0.6. The ephemeris of K2 is known to be better than 1\arcsec, small compared to the beam size. For this reason, and because cometary emission is highly concentrated around the nucleus, we integrated on a single position (i.e.\ 1$\times$1 grid) with beam-switching ($\pm$180 arcsec) sample mode. The observations were taken under good atmospheric conditions with 225 GHz opacity in the range 0.04 to 0.08 (Table \ref{obstable}).  \\

The ACSIS spectrometer was used for all observations, providing a total bandwidth of 250 MHz and a spectral channel spacing of 30.5 kHz, corresponding to 0.04 km s$^{-1}$ at 230 GHz. Pointing and focus were checked roughly every hour using strong sources that were close to the comet at the time of the observation (e.g., CRL\,2688). Flux calibration was done roughly every 3 hours on strong sources (e.g., IRAS\,16293-2422). The data were reduced using \textsc{starlink} and \textsc{IDL}.\\

\begin{figure}
\begin{center}
\hspace{-1 cm}\includegraphics[width=5in,angle=0]{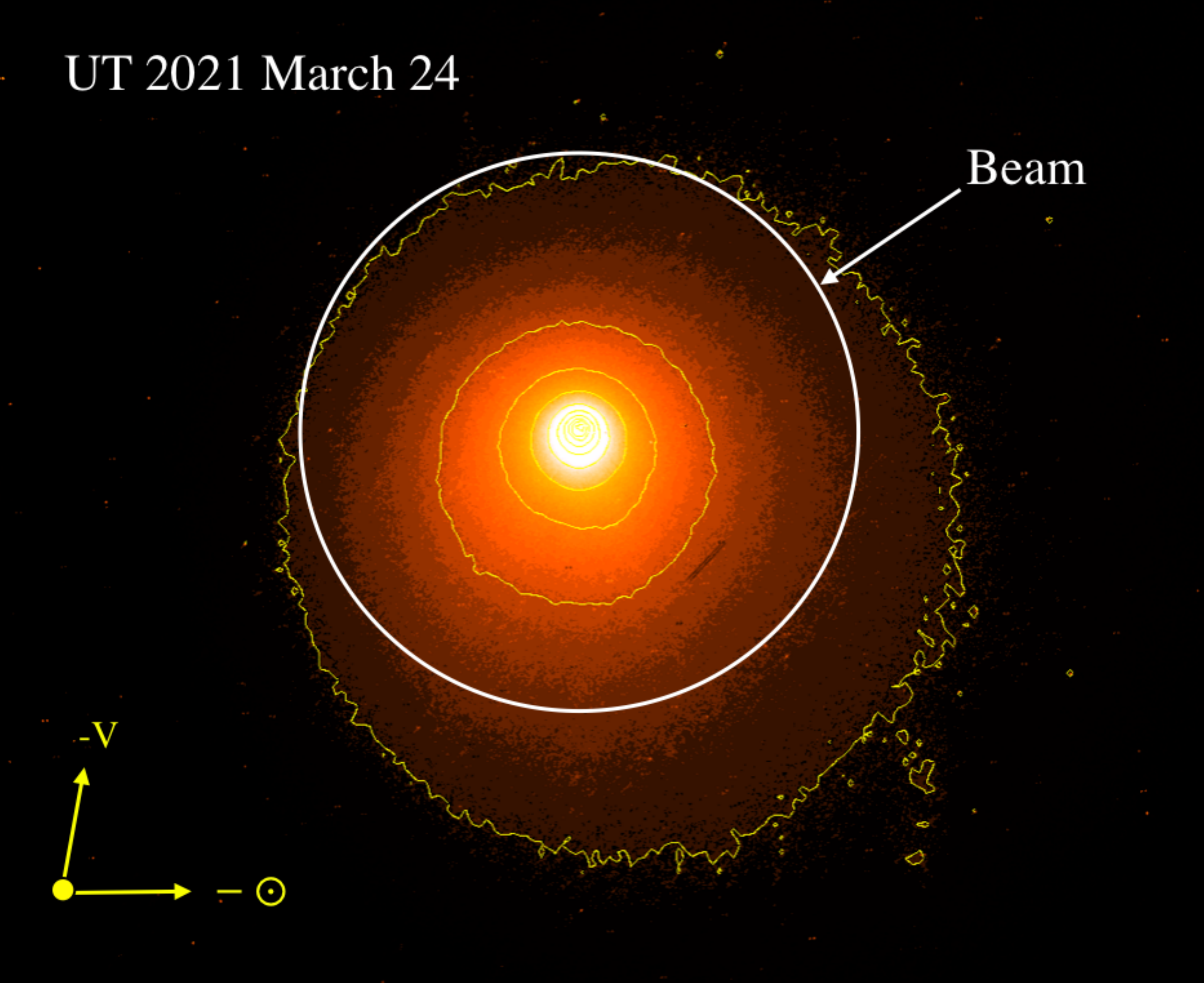}
\caption{Optical image of K2 taken UT 2021 March 24 with the Hubble Space Telescope under program GO 16309. This is the composite of four 200 s integrations taken through the broadband F350LP filter to show the morphology of the cometary dust coma. The 20.41\arcsec~diameter JCMT beam is shown as a white circle.  Direction arrows show the antisolar direction ($-\odot$) and the projected negative heliocentric velocity vector ($-V$). North is to the top, East to the left.}
\label{HST}
\end{center}
\end{figure}

For reference, a Hubble Space Telescope (HST) optical image of K2 taken on UT 2021 March 24 is shown in Figure \ref{HST}, with the size of the JCMT beam indicated by a white circle, corresponding to 1.0$\times 10^5$ km in diameter at the comet. The HST image, from program GO 16309, shows the distribution of dust. There is a strong central concentration (the surface brightness varies inversely with projected distance from the nucleus, consistent with steady-state outflow) with a slight asymmetry caused, in the region shown, by anisotropic ejection from the nucleus. \\

We made correction for the changing geocentric velocity because  the cometary CO lines are very narrow, with widths of just a few $\times$100 m s$^{-1}$. The spectra were accumulated in groups typically with 918s integration each, then averaged after shifting to correct for changes in the geocentric velocity. The resulting composite spectra were then ``baselined" by subtracting linear fits using two symmetric regions, from -10 to -3 km s$^{-1}$ and from +3 to +10 km s$^{-1}$, on each side of the nominal location of the CO J = 2$-$1 line. We experimented with baselines having different spectral windows and found that the choice of the fitted region and the method of the fit were not critical since the baseline deviates only very slightly from linearity in the data. The noise in the data was estimated from the root-mean-square (RMS) value of the main beam brightness temperature fluctuations within $\pm$10 km s$^{-1}$ from the center of the CO line. \\

 \section{Results}
\begin{figure}
\begin{center}
\hspace{-1 cm}\includegraphics[width=5in,angle=0]{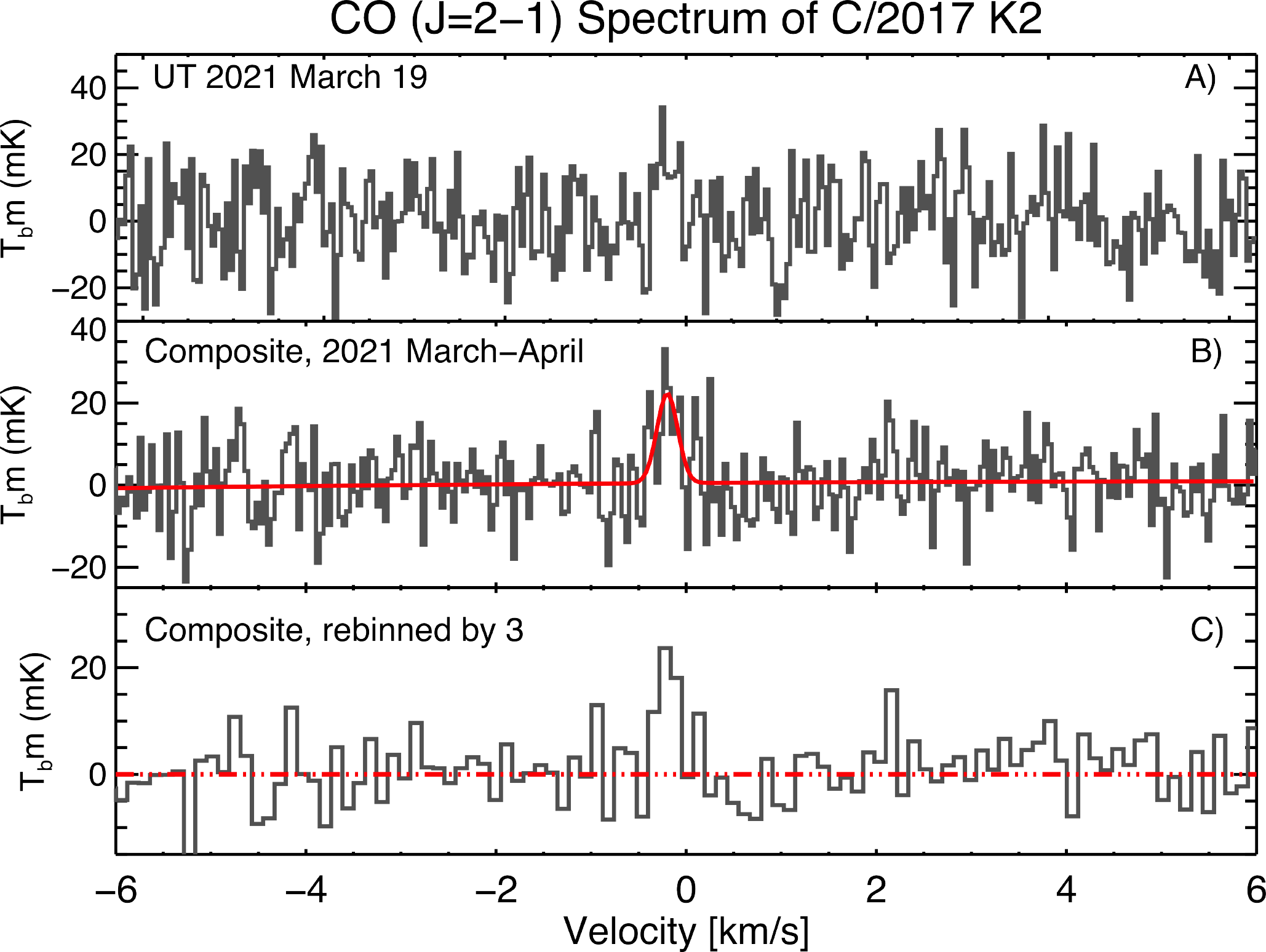}
\caption{Spectrum of emission from the CO (J=2-1) rotational transition obtained from the James Clerk Maxwell Telescope. A) The first detection of the CO line at $\sim$ 2.7$\sigma$, was obtained on UT 2021 March 19 with 1.8 hr integration. B) A composite spectrum taken from UT 2021 March to April (6.62 $\le r_H \le$ 6.83 AU) with 7.7 hr total integration. The spectrum has a velocity resolution of $\sim$0.04 km/s. A Gaussian model fit to the line profile is shown as the solid red line. C) The spectrum shown here is rebinned by 3 spectral channels to a resolution of $\sim$0.12 km/s. The line is slightly blue shifted by 0.20$\pm$ 0.03 km s$^{-1}$ and the fitted full width at half maximum (FWHM) is 0.28 $\pm$ 0.08 km s$^{-1}$. The line area is 8.3 $\pm$ 2.3 mK km s$^{-1}$, corresponding to production rate Q$_{CO} = ($1.6 $\pm$ 0.5) $\times 10^{27}$ s$^{-1}$.}
\label{co_spec}
\end{center}
\end{figure}

\begin{figure}
\begin{center}
\hspace{-1cm}\includegraphics[width=5.5in,angle=0]{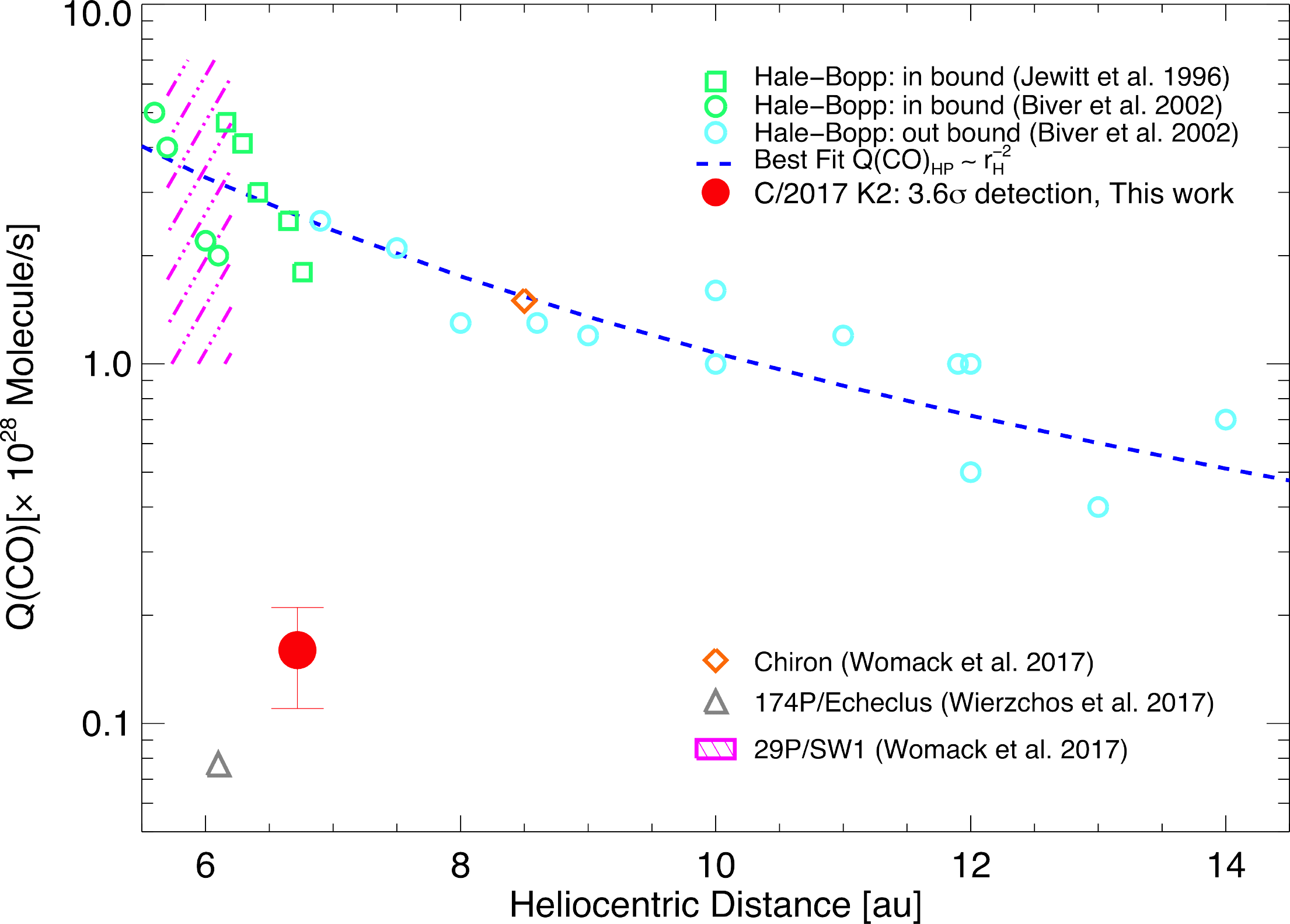}
\end{center}
\caption{CO production rates of objects show distant activity are plotted against heliocentric distance. The observations of C/1995 O1 (Hale-Bopp) follow $Q_{CO}(HB) \propto r_H^{-2}$ \citep{Womack:2017}, shown as a dashed line. A range of CO detections in Centaur 29P/Schwassman-Wachmann is shown by the shaded region, from a summary compilation by \citep{Womack:2017}. The recent detection of CO in active Centaur (60558) 174P/Echeclus is from \cite{Wierzchos:2017}. Outgassing from K2 is about an order of magnitude smaller than the CO production rate of C/1995 O1 (Hale-Bopp) at a comparable distance.}
\label{comparison}
\end{figure}

 The CO(J=2-1) line was initially suspected on UT 2021 March 19 at the 2.7$\sigma$ level when the comet was 6.83 au from the Sun. We obtained and combined additional data from April 1 and 13 (Table \ref{obstable}) to generate a composite spectrum, shown in Figure \ref{co_spec}. Our March observation is so far the furthest detection of CO in an inbound comet, as shown in Figure \ref{comparison}. Using the composite spectrum, we find the CO line has a full width at half maximum (FWHM) of 0.28 $\pm$ 0.08 km s$^{-1}$ (see Figure \ref{co_spec}), consistent with the narrow CO lines detected in other comets \citep{Senay:1994,Jewitt:1996,Biver:1997}. The line area is 8.3 $\pm$ 2.3 mK km/s, corresponding to a 3.6$\sigma$ detection. In addition, the line is slightly blue shifted, with a central velocity at -0.20$\pm$0.03 km/s, likely owing to preferential sublimation of gas in the sunward direction.\\

Gas production rates from comets are estimated using a spatial model of the gas that takes account of the finite lifetimes of the species observed. Often, the Haser model is used \citep{haser:1957}. The solar ionizing flux varies as $r_H^{-2}$, leading to photodestruction lifetimes increasing with distance as $\tau_{CO} = \tau_{0} \cdot r_H^2$, where $\tau_0$=3.6$\times 10^{6}$ s \citep{Huebner:2015} is the lifetime of CO molecule at 1 au. In the case of K2 at 6.72 au, we find $\tau_{CO}$ = 1.6 $\times 10^{8}$ s (about 5 years). This is orders of magnitude longer than the residence time of CO molecules in the beam and, therefore, photodissociation can be neglected. The expected density of CO molecules varies with linear distance from the nucleus as the inverse square, and the projected column density varies with the inverse of the angular distance from the nucleus (as is also true of the dust; \cite{Jewitt:2021}).\\


We write the CO production rate as $Q_{CO}$ [s$^{-1}$] and the gas outflow speed as $V$ [km/s]. Based on the measured line width and assume 10\% thermal broadening effect, we adopted an expansion velocity $V$ = 0.25 km s$^{-1}$, which is consistent with the empirical estimate by the measurements of other comets \citep{Biver:1999,Biver:2000}. The blackbody temperature at this distance is $T_{BB} = $104 K, but we expect that the temperature of the CO sublimating surface should be strongly depressed by the loss of energy to the destruction of bonds in the CO ice, for which the latent heat of sublimation is only $H = 2\times10^5$ J kg$^{-1}$ \citep{Huebner:2006}. We solved the energy balance equation for a flat and perfectly absorbing CO ice surface oriented perpendicular to the Sun-comet line and located at $r_H$ = 6.72 au, finding temperature depression to $T =$ 26 K. Even lower rotational temperatures were measured in comets 29P/Schwassmann-Wachmann 1 near $r_H$ =6 au \citep[T $\sim$10 K,][]{Crovisier:1995} and in C/1995 O1 (Hale-Bopp) near $r_H$ = 10 au \citep[T = 8 K, ][]{Gunnarsson:2003}, where they result from adiabatic cooling  as the gas expands away from the nucleus. With only a single rotational line, we have no independent measure of the temperature in K2, but we confidently assume that comparably low temperatures prevail in the coma of K2. Following the method described in \cite{Drahus:2010}, we assume local thermodynamic equilibrium (LTE) conditions are present throughout the coma and the photodissociation effects are negligible. With $T$ = 26 K, we derived a model dependent production rate of $Q_{CO} = (1.6\pm0.5) \times10^{27}$ s$^{-1}$. \\

\section{Discussion}

The equilibrium CO sublimation rate for a flat and perfectly absorbing  surface oriented perpendicular to the Sun-comet line at $r_H$ = 6.7 au is $f_{CO} = 1.0\times10^{-4}$ kg m$^{-2}$ s$^{-1}$. To supply $Q_{CO}$ requires an exposed CO ice patch of area  

\begin{equation}
A = \mu m_H \frac{ Q_{CO}}{f_{CO}}
\end{equation}
    
\noindent where $\mu$ = 28 is the molecular weight of CO and $m_H = 1.67\times10^{-27}$ kg is the mass of hydrogen. Substituting $Q_{CO} =1.6\times10^{27}$ s$^{-1}$, we find $A = 7.5\times10^5$ m$^2$, corresponding to a circular patch only 0.5 km in radius. \\

High resolution imaging observations set only an upper limit to the radius of the nucleus, $r_n \le$ 9 km \citep{Jewitt:2017}. Given this limiting radius, the effective fraction of the surface of the nucleus that is actively sublimating CO is a modest $f_A = A/(4\pi r_n^2) \ge 7\times10^{-4}$. This value may be compared to the active fractions of short-period comets, which occupy a wide range but which are commonly $\sim$1\% \citep{A'Hearn:1995}. It is additionally possible that part of the CO is released from the dust grains in the coma. Indeed, these grains strongly dominate the scattering cross-section of the comet relative to the of the nucleus, so some coma contribution might be expected.  \\

We estimate the mass of CO lost from K2 since activity began (at $r_H \sim$ 35 AU) from 

\begin{equation}
\Delta M_{CO} = \mu m_H \int_{t_1}^{t_2} Q_{CO}(r_H(t)) dt
\end{equation}

\noindent where $t_1$ and $t_2$ are the start and end times and $r_H(t)$ is from the JPL Horizons ephemeris. Substituting $Q_{CO} =(1.6\pm0.5)\times10^{27} (6.7/r_H(t))^2$, we obtain $\Delta M_{CO} = (6\pm2)\times10^9$ kg, corresponding to an ice thickness $\Delta r_n = \Delta M_{CO}/(\rho A)$ = 15$\pm$5 m, where $\rho$ = 500 kg m$^{-3}$ is the assumed density of the nucleus. In this view, sublimation from a 1 km$^2$ exposed CO patch would have eroded a pit only a few meters deep. Sublimation could also occur from a larger area, excavated to a smaller depth, or from a different geometry. For example, some in-situ observations of active short-period comet nuclei suggest that sublimation proceeds by cliff erosion, not just from flat-bottomed pits \citep{Keller:2015}.\\

The CO production rate from K2 is about 10\% of that from comet C/1995 O1 (Hale-Bopp) when at similar heliocentric distance (Figure \ref{comparison}). If these two nuclei have the same active fractions, then the production rate ratio would indicate nucleus radii in the ratio $r_n(K2)/r_n(HB) = 10^{-1/2} \sim 0.3$. Estimates for the nucleus radius in C/1995 O1 (Hale-Bopp) vary over the range 13 $\le r_N(HB) \le$ 50 km \citep{Weaver:1997, Lamy:1999}, suggesting that 4 $\le r_N(K2) \le$ 15 km. This range overlaps the limit, $r_N(K2) \le$ 9 km, set from the optical profile of K2 \citep{Jewitt:2017} but does not improve it. In C/1995 O1 (Hale-Bopp), the CO production varied as $r_H^{-2}$ \citep{Womack:2017}. If this variation holds in K2, we expect that the perihelion (1.80 AU) production rate will reach $\sim2.2\times10^{28}$ s$^{-1}$ ($\sim 10^{3}$ kg s$^{-1}$). \\

The gas mass production rate in CO is $dM_{CO}/dt = \mu m_H Q_{CO}$ which, by substitution, gives $dM_{CO}/dt$ = 75$\pm$23 kg s$^{-1}$. This compares with the dust mass production rate inferred from measurements of the optical continuum, $dM_d/dt = 10^3 a_1^{1/2} (10/r_H)^{2}$, with $a_1$ being the mean dust particle radius expressed in millimeters. Morphological considerations show that $a_1$ = 0.1 to 1 \citep{Jewitt:2019}. At $r_H$ = 6.7 au, we find $dM_d/dt$ = 700 to 2200 kg s$^{-1}$. Therefore, it is likely that the dust/gas ratio in K2 substantially exceeds unity.  This is true of many or most comets (e.g.~ratios up to $\sim$30 have been reported in 2P/Encke; \cite{Reach2000}), and is possible because although most of the mass is contained in solid grains, these grains are expelled slowly compared to the gas flow speed and the momentum is always dominated by the gas. \\

\section{Summary}
1) We used the James Clerk Maxwell Telescope to detect the J = 2$-$1 rotational transition of carbon monoxide in C/2017 K2 when inbound at heliocentric distance $r_H = 6.72$ au. The line is blue-shifted by 0.20$\pm$0.03 km s$^{-1}$ with area and width 8.3$\pm$2.3 mK km s$^{-1}$ and 0.28$\pm$0.08 km s$^{-1}$, respectively. \\

2) The corresponding carbon monoxide production rate is $Q_{CO} = (1.6\pm0.5) \times10^{27}$ s$^{-1}$ ($dM_{CO}/dt$ = 75$\pm$23 kg s$^{-1}$). This value is smaller than the production rate inferred from measurements of coma dust, indicating the dust/gas ratio $>$ 1. \\

3) The measured CO production can be supplied by surface sublimation of CO ice from a modest (1 km$^2$) surface patch. We estimate that $\sim10^9$ kg of CO has already been lost of the nucleus of C/2017 K2.\\

\section{Acknowledgment}
We thank the JCMT staff for their assistance. This work is based in part on observations made with the NASA/ESA Hubble Space Telescope, obtained from the archive at the Space Telescope Science Institute (STScI). STScI is operated by the association of Universities for Research in Astronomy, Inc.\ under NASA contract NAS 5-26555. These observations are associated with GO program 16309.  Y.Z. thanks the National Natural Science Foundation of China (Grant Nos. 12073084, 11761131008). This work makes use of the JCMT data obtained under Program ID: S21AP001. The JCMT is operated by the EAO on behalf of NAOJ; ASIAA; KASI; CAMS as well as the National Key R\&D Program of China (No. 2017YFA0402700). Additional funding support is provided by the STFC and participating universities in the UK and Canada.


\newpage
\begin{deluxetable}{lccccc}\tablewidth{2.3in}
\tabletypesize{\scriptsize}
\tablecaption{ Observational Parameters 
  \label{obstable}}
\tablecolumns{6} \tablehead{  \colhead{UT Date }  &\colhead{t$_{int}$ \tablenotemark{a}} &  \colhead{$r_H$ \tablenotemark{b}}  & \colhead{$\Delta$ \tablenotemark{c}} &   \colhead{ $\alpha$ \tablenotemark{d}}  &  \colhead{$\tau_{225}$ \tablenotemark{e}}  \\
\colhead{}&\colhead{(s)} &\colhead{[au]}&\colhead{[au]}&\colhead{[deg]}&\colhead{}}
\startdata
2021 April 13 & 12852 & 6.62 & 6.45 & 8.66 & 0.04 - 0.06 \\
2021 April 01 & 8262 & 6.72 & 6.65 & 8.54 & 0.05 - 0.07  \\
2021 March 19 & 6426 & 6.83 & 6.85 & 8.34 &  0.06 - 0.08 \\
\enddata
\tablenotetext{a}{Total integration time in seconds.}
\tablenotetext{b}{Average heliocentric distance.}
\tablenotetext{c}{Average geocentric distance.}
\tablenotetext{d}{Solar phase angle (Sun-object-Earth).}
\tablenotetext{e}{Atmospheric optical depth at 225 GHz.}
\end{deluxetable}

\clearpage

\clearpage

\clearpage

\end{document}